  \documentstyle[preprint,prl,aps]{revtex}
\begin{document}
\draft
\title{Nearby Doorways, Parity Doublets and Parity Mixing in Compound
Nuclear States}
\author{N. Auerbach,$^{(1)}$ J.D. Bowman$^{(2)}$ and V. Spevak$^{(1)}$}
\address{$^{(1)}$School of Physics and Astronomy, Tel Aviv University,
Tel Aviv 69978, Israel}
\address{$^{(2)}$Los Alamos National Laboratory, Los Alamos,
New Mexico 87545}
\date{August 17, 1994}
\maketitle
\begin{abstract}
We discuss the implications of a doorway state model for parity mixing
in compound nuclear states. We argue that in order to explain the
tendency of parity violating asymmetries measured in $^{233}$Th to
have a common sign, doorways that contribute to parity mixing must be
found in the same energy neighbourhood of the measured resonance. The
mechanism of parity mixing in this case of nearby doorways is closely
related to the intermediate structure observed in nuclear reactions in
which compound states are excited. We note that in the region of
interest ($^{233}$Th) nuclei exhibit octupole deformations which leads
to the existence of nearby parity doublets. These parity doublets are
then used as doorways in a model for parity mixing. The contribution
of such mechanism is estimated in a simple model.
\end{abstract}

\narrowtext

Recent experiments on parity violation in compound nuclear (CN) states
\cite{Bowman1,Bowman2} are providing new information on the parity
non-conserving (PNC) interaction. The experimental results for
$^{232}$Th showed that measured PNC asymmetries fluctuated about
non-zero average, in contradiction to the purely random behavior
expected on the basis of the statistical model of the CN. The
understanding of PNC phenomena challenges theory to treat
simultaneously the chaotic and the regular aspects of the CN system. A
number of attempts were made to explain the new results in the
framework of statistical models \cite{stat-stuff} or in models that
combine nuclear dynamics with the statistical aspects
\cite{AuR92,Bow-John}. Among the latter the doorway state approach was
used in several theoretical works \cite{AuR92,Bow-John}. In Ref.\
\cite{AuR92} the spin-dipole (SD) resonance was used as the doorway to
describe the PNC spreading width of the compound resonances. Later the
same model was applied \cite{Auer-Bow} in the calculation of the
average longitudinal asymmetry. In Ref.\ \cite{Bow-John} the $s_{1/2}$
and $p_{1/2}$ single-particle states were used as the doorways in an
attempt to explain the constancy of signs. The common feature of these
models is that they deal with distant doorways and involve only the
one-body part of the  PNC interaction. The term ``distant doorways''
refers to the fact that the position of the doorways is removed by
\mbox{$1\hbar\omega \sim 7$~MeV} from the CN resonances under
consideration.  The model in Ref.\ \cite{AuR92} which involved the
collective effects of the spin-dipole resonance was able to account
for the PNC spreading width when a reasonable \cite{DDH} value for the
one-body PNC matrix element was used. However, when the same size
matrix element was used the average asymmetry was almost two orders of
magnitude smaller than measured. Recently more extensive
theoretical investigations characterized several terms that could
contribute to PNC asymmetries in the CN
\cite{Lew-Weid,Auerbach-Spevak}. Additional terms were derived and
evaluated. However these theories did not identify new terms that were
large enough to explain the large non-fluctuating asymmetry.

The quantity measured in the experiments with polarized epithermal
neutrons is the longitudinal asymmetry
\begin{equation}
P(E_{r})=\frac{\sigma^{^{+}}(E_{r})-\sigma^{^{-}}(E_{r})}
{\sigma^{^{+}}(E_{r})+\sigma^{^{-}}(E_{r})} \ ,
\label{1}
\end{equation}
where $\sigma^{^{+}}$, $\sigma^{^{-}}$ are the resonance part of the
total cross sections for neutrons with positive and negative
helicities respectively. The index $r$ designates the fact that the
scattering is to a compound resonance $\vert r\rangle$ at energy
$E_{r}$, carrying the quantum numbers \mbox{$J=\frac{1}{2}^{-}$.} We
will refer to these resonances as $p_{1/2}$. The leading term of the
asymmetry P can be written \cite{Auer-Bow,Auerbach-Spevak}
\begin{equation}
P(E_{r})=-2\sum_{s}\frac
{\langle s\vert V^{\text{PNC}}\vert r\rangle}{E_{r}-E_{s}}
\frac{\gamma_{s}}{\gamma_{r}} \ ,
\label{2n}
\end{equation}
where $\gamma_{r}$ and $\gamma_{s}$ denote the escape amplitudes from
resonances $\vert r\rangle$ and $\vert s\rangle$ due to the strong
interaction force and are given by:
\begin{eqnarray}
\gamma_{r}=\langle \Phi_{p}^{^{(+)}}\vert H\vert r\rangle \ ,\nonumber \\
\gamma_{s}=\langle \Phi_{s}^{^{(-)}}\vert H\vert s\rangle \ ,
\label{3n}
\end{eqnarray}
where $\Phi_{p}^{^{(+)}}$ and $\Phi_{s}^{^{(-)}}$ denote the continuum
$p$ and $s$-wave functions in the elastic channels. The sum in Eq.\
(\ref{2n}) extends in principle over all states $\vert s\rangle$ that
have the quantum numbers \mbox{$J=\frac{1}{2}^{+}$.} To reduce the sum
in the equation one often introduces the doorway state approximation
\cite{Auerbach72,Feshbach-Lemmer}. One seeks a subspace of states
$\vert s\rangle$, denoted by $\vert d\rangle$ such that the coupling
between these states and the states $\vert r\rangle$ --
\mbox{$\langle r\vert V^{\text{PNC}}\vert d\rangle$}
is sizeable or that the coupling between $\vert d\rangle$ and the
continuum is strong or that both conditions are fulfilled so that when
the sum in Eq.\ (\ref{2n}) is replaced by the partial sum over states
$\vert d\rangle$ the result will be a good approximation to Eq.\
(\ref{2n}). In a  formal treatment
\cite{Auerbach-Spevak,Auerbach72,Kerman-deTol} one divides the space
of states $\vert s\rangle$ into: \mbox{$\{ s\} =\{ d\} +\{ s'\}$.}
Assuming that the signs of the matrix elements
\mbox{$\langle r\vert V^{\text{PNC}}\vert s'\rangle$} are randomly
distributed one derives
\cite{Auerbach-Spevak,Auerbach72,Feshbach-Lemmer,Kerman-deTol} the
following expression:
\begin{equation}
P(E_{r})\simeq -2\sum_{d}\frac{(E_{r}-E_{d})
\langle d\vert V^{\text{PNC}}\vert r\rangle \gamma_{d}}
{[(E_{r}-E_{d})^{2}+\frac{\Gamma_{d}^{2}}{4}]\gamma_{r}} \ ,
\label{9}
\end{equation}
where $\gamma_{d}$ is the escape amplitude of the doorway. Here
$\Gamma_{d}$ is the width of the doorway resulting from the coupling
of $\vert d\rangle$ to the states $\vert s'\rangle$. The PNC matrix
element is purely imaginary and the product of the two escape
amplitudes is nearly pure imaginary because it involves opposite
parities and thus the product:
\mbox{$(\langle d\vert V^{\text{PNC}}\vert r\rangle\gamma_{r}
\gamma_{d})$}
is nearly real.

The PNC spreading width in the doorway approximation is given by
\cite{AuR92}:
\begin{equation}
\Gamma_{r}^{\downarrow \text{PNC}}
=\sum_{d} \frac{\vert\langle r\vert V^{\text{PNC}}
\vert d\rangle\vert^{2}}
{(E_{d}-E_{r})^{2}+\frac{\Gamma_{d}^{2}}{4}}
\Gamma_{d}^{\downarrow} \ ,
\label{11}
\end{equation}
where $\Gamma_{d}^{\downarrow}$ is the spreading width of the doorway.

In the analysis of the recent class of experiments two basic
quantities are determined \cite{Bowman1,Bowman2}.
Taking the ensemble of states
$\vert r\rangle$ one determines the average asymmetry $\overline{P}$;
the fluctuating part of the measured asymmetries yields the PNC
spreading width in a given nucleus. In $^{233}$Th the analysis
\cite{Bowman2} yields for
\mbox{$\Gamma_{r}^{\downarrow \text{PNC}}\simeq 7.4\times 10^{-7}$~eV}
and \mbox{$\overline{P}\simeq (8 \pm 6)\%$.}
Equations (\ref{9}) and (\ref{11}) will be the basis of our discussion
of the above two quantities.

Let us first apply these equations to the case of ``distant''
doorways. This was already the subject of study in Refs.\
\cite{Bow-John} and \cite{Auerbach-Spevak} and we repeat this case in
order to emphasize the difficulty one faces. Consider a one-body PNC
potential of the form \cite{Michel}
\begin{equation}
V^{\text{PNC}}=\epsilon 10^{-7}\frac{1}{2}\sum_{i}
\{ f(r_{i}),\mbox{\boldmath $\sigma$}_{i}\mbox{\boldmath $p$}_{i}c\} \ ,
\label{12}
\end{equation}
where $f(r)$ is some function of the distance $r$ and the curly
brackets denote the anticommutation operation. For this potential the
doorway that will couple strongly to the state $\vert r\rangle$ is the
giant spin-dipole  resonance built on the state $\vert r\rangle$ i.e.\
by acting with  the operator \mbox{\boldmath$\sigma r$} on $\vert
r\rangle$. (For more details see Refs.\ \cite{AuR92,Auerbach-Spevak}).
The energy centroid of the spin-dipole resonance built on
$\vert r\rangle$ is in $^{233}$Th about 7~MeV above $E_{0}$. The width
of the spin-dipole is several MeV, and for the sake of an estimate let
us take \mbox{$\Gamma_{\text{SD}}^{\downarrow}=3$~MeV.}
In Ref.\ \cite{Auer-Bow,Auerbach-Spevak} the ratio of the escape
amplitudes in Eq.\ (\ref{9}) was shown to be:
\begin{equation}
\frac{\gamma_{_{\text{SD}}}}{\gamma_{r}}\simeq \frac{1}{\sqrt{N}}
\frac{\sqrt{3}}{kR}\simeq \frac{1}{\sqrt{N}}\cdot 10^{3} \ ,
\label{13}
\end{equation}
where $\frac{\sqrt{3}}{kR}$ is the ratio of penetrabilities and
\mbox{$N\sim 100$} is the number of particle-hole excitations that
make up the collective spin-dipole in $^{232}$Th.
Using Eq.\ (\ref{9}) and the value for $\overline{P}$ of $8\%$ and the
above values for the width and energy of the SD-doorway we find:
\begin{equation}
\langle r\vert V^{\text{PNC}}\vert \text{SD}\rangle =1700~\mbox{eV} \ .
\label{14}
\end{equation}
If we use now Eq.\ (\ref{11}) to estimate the collective PNC matrix
element with the above values for $E_{d}$ and
$\Gamma_{d}^{\downarrow}$, we find that
\mbox{$\langle r\vert V^{\text{PNC}}\vert \text{SD}\rangle =3$~eV.}
There is a discrepancy of almost three orders of magnitude with the
value obtained in Eq.\ (\ref{14}). The value of $3$~eV is quite
consistent with the predictions of the existing models for the PNC
interaction in nuclei \cite{DDH}.

This leads us to the conclusion that one must go beyond the one-body
aspects of the theory and the idea of distant doorways and address the
question of the role of nearby doorways.

Let us consider the following physical scenario: the compound state
which shows up in experiment as parity mixed, is close to a doorway so
that the the difference \mbox{$\vert E_{r}-E_{d}\vert <\Gamma_{d}$}
and that \mbox{$\Gamma_{d}\simeq 100$~eV.} Let us take for the sake of
the estimate the value of
\mbox{$\vert E_{r}-E_{d}\vert =\Gamma_{d}/ 2$.} Let us also assume
that the ratio $\gamma_{d}/\gamma_{r}$ is as before $10^{3}$ due to
the penetrability effect. Using again Eq.\ (\ref{9}) we find for
the PNC matrix element the value
\mbox{$\langle r\vert V^{\text{PNC}}\vert d\rangle =4
\times 10^{-3}$~eV}
while when using Eq.\ (\ref{11}) we estimate for this matrix element
the value
\mbox{$\langle r\vert V^{\text{PNC}}\vert d\rangle =6
\times 10^{-3}$~eV.}
The two numbers are not inconsistent as they were in the previous
``distant'' doorway case. Of course this consistency does not serve as
a prove, it is just an indication that the discrepancy can be resolved
when there are nearby doorways that couple to the compound state
through a PNC interaction.

Let us now develop this idea further. In various reactions with low
energy protons or neutrons and with good resolution one observes in
the excitation functions structures that have a width $\Gamma_{d}$
that is intermediate between the single-particle $\Gamma_{s.p.}$ and
the compound width $\Gamma_{r}$,
\mbox{$\Gamma_{s.p.}\gg \Gamma_{d}\gg \Gamma_{r}$}
\cite{Feshbach-Lemmer,Mekjian}. These intermediate resonances are
usually non-overlapping and their spacings $D_{d}$ are such that
\mbox{$D_{s.p.}\gg D_{d}>D_{r}$.}
Very fine resolution experiments show that the intermediate resonances
are composed of a number of compound resonances and that the
intermediate resonances are enhancements in the excitation function of
the compound nuclear states.

In heavy nuclei in the vicinity of \mbox{$A=240$} intermediate
structure was seen and studied extensively. The most striking and most
intriguing example is the intermediate structure observed in neutron
induced fission $(n,f)$. The cross section seen in these reactions
shows groupings of compound states that are enhanced. The width of
each such enhanced bump is about \mbox{$\Gamma_{d}=200$~eV} and the
bumps are spaced about \mbox{$D_{d}=700$~eV.} A very elegant
explanation of this behaviour of the $(n,f)$ excitation function was
given in terms of a double hump fission barrier\cite{Bjorn-Lynn}. The
intermediate resonances observed in the reaction are due to states in
the second more shallow and lower fission barrier from which fission
is easier and therefore enhanced.

Let us now use this information about the intermediate structure to
consider some models for parity violation in these heavy nuclei. Let
us first consider a model which connects to the observation of
intermediate structure in the $(n,f)$ reaction. The transition
amplitude  for this process can be expressed in the following form:
\begin{equation}
T_{fn}=\sum_{q^{+}}\frac{\langle f^{+}\vert H
\vert q^{+}\rangle\langle q^{+}\vert H\vert\Phi_{n}^{^{(+)}}\rangle}
{E-E_{q^{+}}+\frac{i}{2}\Gamma_{q^{+}}} \ ,
\label{17}
\end{equation}
where by $\vert f^{+}\rangle$ we denoted a state of positive parity
describing the fissioning process, $\vert\Phi_{n}^{^{(+)}}\rangle$ the
incoming neutron continuum wave function states and
$\vert q^{+}\rangle$ are positive parity compound states. The fact
that the $(n,f)$ process shows intermediate structure at some
excitation energies, while the $(n,n)$ excitation function does not,
indicates that the mixing is enhanced for some states
$\vert q^{+}\rangle$ that are in the vicinity of the fissioning states
$\vert f^{+}\rangle$.

The two-body effective PNC interaction, derived using the
meson-exchange models \cite{DDH} and the two-body nuclear force  have
both similar ranges. Let us therefore assume that the matrix
elements \mbox{$\langle f^{+}\vert H\vert q^{+}\rangle$}
and \mbox{$\langle r\vert V^{\text{PNC}}\vert f^{+}\rangle$}
are related.

One can write
\begin{equation}
T=\sum_{r}
\frac{\langle\Phi_{s}^{^{(-)}}\vert H\vert f^{+}\rangle
\langle f^{+}\vert V^{\text{PNC}}\vert r\rangle \langle r\vert H\vert
\Phi_{p}^{^{(+)}}\rangle}{(E-E_{r}+\frac{i}{2}\Gamma_{r})
(E-E_{f^{+}}+\frac{i}{2}\Gamma_{f^{+}})} \ ,
\label{18}
\end{equation}
which is the PNC $T$ matrix for neutron scattering in the doorway
state approximation with the doorway being the fission state $\vert
f^{+}\rangle$. The corresponding asymmetry is then
\begin{equation}
P(E_{r})=-2\frac{
(E_{r}-E_{f^{+}})
\langle\Phi_{s}^{^{(-)}}
\vert H\vert f^{+}\rangle\langle f^{+}\vert V^{\text{PNC}}
\vert r\rangle}
{\langle \Phi_{p}^{^{(+)}} \vert H\vert r\rangle
[(E_{r}-E_{f^{+}})^{2}+\frac{1}{4}\Gamma_{f^{+}}^{2}]} \ .
\label{19}
\end{equation}
The enhancements observed in the $(n,f)$ reaction will also occur in
the asymmetry given in Eq.\ (\ref{19}) if the compound state
$\vert r\rangle$ is in the vicinity of the doorway
$\vert f^{+}\rangle$.

It is remarkable that in $^{233}$Th (as well as in $^{231}$Th) one
finds \cite{Bjorn-Lynn,Moller-Nix,Blons84+89} that the energy
potential as a function of deformation is triple humped. It has been
shown \cite{Moller-Nix} that the existence of the third minimum in the
potential is related to the appearance of {\em pear-like octupole
deformations} in the body frame of reference. This spontaneous
breaking of reflection symmetry implies the existence of nearby (a few
tens of keV apart) parity doublets
\cite{Bohr-Mottelson,Ahmad-Butler+Aberg-rev}. In fact at somewhat
higher energies than considered here such doublets were observed in
$^{233}$Th in the $(n,f)$ experiments \cite{Blons84+89}.
It is worth mentioning that the states in the third minima with
excitation energies \mbox{$E_{\text{exc}}\sim 4-8$~MeV} belong to the class
of hyperdeformed states, since the quadrupole and octupole
deformations are very large, of the order \mbox{$\beta_{2}\simeq
0.90$,} \mbox{$\beta_{3}\simeq 0.35$} \cite{NazPLB+in,Moller-Nix}.

Consider now such a \mbox{$J=\frac{1}{2}^{\pm}$} doublet denoted by
$\vert f^{\pm}\rangle$. The energy difference
\mbox{$\vert E_{f^{+}}-E_{f^{-}}\vert$}
is equal to a few tens of keV, which is of course small but
larger than \mbox{$\vert E_{r}-E_{f^{+}}\vert$} or $\Gamma_{f^{+}}$.
Let us now single out the \mbox{$\vert f^{-}\rangle$} component in the
wave function of $\vert r\rangle$
\begin{equation}
\vert r\rangle=a_{rf^{-}}\vert f^{-}\rangle+\vert r' \rangle \ .
\label{in1}
\end{equation}
Then
\begin{equation}
\langle f^{+}\vert V^{\text{PNC}}\vert r\rangle = a_{rf^{-}}
\langle f^{+}\vert V^{\text{PNC}}\vert f^{-}\rangle \ .
\label{in2}
\end{equation}
We drop the contribution of $\vert r'\rangle$ to the matrix element in
the above equation because: ({\em i}) in view of the $(n,f)$
experimental results the coupling of $\vert r\rangle$ to $\vert
f^{-}\rangle$ should dominate, meaning that $a_{rf^{-}}$ is relatively
large and ({\em ii}) because of the close relation between $\vert
f^{+}\rangle$ and $\vert f^{-}\rangle$ the PNC matrix element should
also be relatively large. Substituting Eq.\ (\ref{in2}) into Eq.\
(\ref{19}) we find:
\begin{eqnarray}
&&P(E_{r}) \nonumber \\
&&=-2 \frac{\langle f^{+}\vert V^{\text{PNC}}\vert f^{-}\rangle
 (E_{r}-E_{f^{+}})
\langle\Phi_{s}^{(-)}\vert H\vert f^{+}\rangle a_{rf^{-}}}
{[(E_{r}-E_{f^{+}})^{2}+\frac{1}{4}\Gamma_{f^{+}}^{2}]
\langle \Phi_{p}^{(+)}\vert H\vert r\rangle} \ . \nonumber \\
\label{in3}
\end{eqnarray}
In order to evaluate the average part of $P(E_{r})$ let us proceed
with this model and use Eq.\ (\ref{in1}) to estimate
\mbox{$\langle \Phi_{p}^{(+)}\vert H\vert r\rangle $}.
Recognizing as before that the component of $\vert f^{-}\rangle$ in
$\vert r\rangle$ is relatively large and that $\vert f^{-}\rangle$
connects to exit channel as the $(n,f)$ experiments show we may write
\begin{equation}
\langle \Phi_{p}^{(+)}\vert H\vert r\rangle \simeq a_{rf^{-}}
\langle \Phi_{p}^{(+)}\vert H\vert f^{-}\rangle \ .
\label{in4}
\end{equation}
Substituting this into Eq.\ (\ref{in3}) we find:
\begin{equation}
P(E_{r})=-2 \frac{
\langle f^{+}\vert V^{\text{PNC}}\vert f^{-}\rangle
(E_{r}-E_{f^{+}})
\langle\Phi_{s}^{(-)}\vert H\vert f^{+}\rangle}
{[(E_{r}-E_{f^{+}})^{2}+\frac{1}{4}\Gamma_{f^{+}}^{2}]
\langle \Phi_{p}^{(+)}\vert H\vert f^{-}\rangle} \ .
\label{in5}
\end{equation}
We see that except for the energy difference in the numerator the sign
of $P$ is independent of \mbox{$\vert r\rangle$} and is fixed. If the
compound states \mbox{$\vert r\rangle$} observed in experiment are
located on one side of $E_{f^{+}}$ the asymmetries $P$ will have the
same sign. A change in sign should occur when the energies $E_{r}$
cross $E_{f^{+}}$. This is similar to the circumstances that occur in
isobaric analog resonances \cite{Feshbach-Lemmer,Mekjian}.

Realizing the fact that we deal with parity doublets having similar
structure we take the ratio of the two escape amplitudes from
$\vert f^{+}\rangle$ and $\vert f^{-}\rangle$ to depend only on the
penetrability factor which is about $10^{3}$. Taking from the
experiment \cite{Bowman2} the value
\mbox{$\overline{P}\simeq 8\times 10^{-2}$} and using
\mbox{$\vert E_{r}-E_{f^{+}}\vert =\Gamma_{f^{+}}/ 2$} we derive from
Eq.\ (\ref{in5}):
\begin{equation}
\frac{\langle f^{+}\vert V^{\text{PNC}}\vert f^{-}\rangle}
{\Gamma_{f^{+}}} \simeq 4\times 10^{-5} \ .
\label{in6}
\end{equation}
For \mbox{$\Gamma_{f^{+}} \simeq 200$~eV} the PNC matrix element is
\mbox{$\vert\langle f^{+}\vert V^{\text{PNC}}\vert f^{-}\rangle\vert
\simeq 10^{-2}$~eV}.
This is not an unreasonable value for such matrix element. Recall that
in Ref.\ \cite{AuR92,Auer-Bow} we estimate that single-particle
configurations give PNC matrix elements of about $0.3$~eV.

We should mention that the breaking of reflection symmetry in the body
frame of reference in actinides \mbox{$A\leq 229$} is well
established. These nuclei exhibit \cite{Leander-Nazarewicz-Bertsch}
(see reviews \cite{Ahmad-Butler+Aberg-rev}) at low-energies octupole
deformations \mbox{$\beta_{3}\simeq 0-0.12$,} enhanced E1 transitions
and large dipole moments in the body frame of reference.

We also note that effects of parity violation were observed
in fission reactions induced by polarized neutrons
\cite{Petrov89+94,Sushkov-Flamb-f,Bunakov-Gudkov}.
A theoretical discussion of this phenomenon was presented some years
ago \cite{Sushkov-Flamb-f} in which the notion of octupole
deformations and parity doublets was invoked. The connection between
PNC effects in the $(n,f)$ reactions and the PNC experiments discussed
in the present work should therefore not be overlooked.

We wish to thank G.F.~Bertsch, V.V.~Flambaum, G.T.~Garvey,
W.C.~Haxton, M.B.~Johnson, A.K.~Kerman and D.H.~Wilkinson
for helpful discussions.
This work was supported in part by the US-Israel Binational Science
Foundation. Part of this work was performed at the Institute for
Nuclear Theory in Seattle.

\end{document}